\documentclass[sigconf]{acmart}
\usepackage{tabularx}
\usepackage{array}
\usepackage{booktabs}
\usepackage{multirow}
\AtBeginDocument{%
  }

\copyrightyear{2025}
\acmYear{2025}
\setcopyright{cc}
\setcctype{by}
\acmConference[DIS '25]{Designing Interactive Systems Conference}{July 5--9, 2025}{Funchal, Portugal}
\acmBooktitle{Designing Interactive Systems Conference (DIS '25), July 5--9, 2025, Funchal, Portugal}\acmDOI{10.1145/3715336.3735831}
\acmISBN{979-8-4007-1485-6/2025/07}

\begin{document}

\title[Improving Public Service Chatbot Design and Civic Impact]{Improving Public Service Chatbot Design and Civic Impact: Investigation of Citizens' Perceptions of a Metro City 311 Chatbot}

\author{Jieyu Zhou}
\affiliation{%
  \institution{Georgia Institute of Technology}
  \city{Atlanta}
  \state{Georgia}
  \country{USA}
}
\email{jzhou625@gatech.edu}

\author{Rui Shen}
\affiliation{%
  \institution{Georgia Institute of Technology}
  \city{Atlanta}
  \state{Georgia}
  \country{USA}
}
\email{rshen65@gatech.edu}

\author{You Yue}
\affiliation{%
  \institution{Georgia Institute of Technology}
  \city{Atlanta}
  \state{Georgia}
  \country{USA}
}
\email{yueyou@gatech.edu}

\author{Carl DiSalvo}
\affiliation{%
  \institution{Georgia Institute of Technology}
  \city{Atlanta}
  \state{Georgia}
  \country{USA}
}
\email{cdisalvo@gatech.edu}

\author{Lynn Dombrowski}
\affiliation{%
  \institution{Georgia Institute of Technology}
  \city{Atlanta}
  \state{Georgia}
  \country{USA}
}
\email{goblyn@gatech.edu}

\author{Christopher J. MacLellan}
\affiliation{%
  \institution{Georgia Institute of Technology}
  \city{Atlanta}
  \state{Georgia}
  \country{USA}
}
\email{cmaclell@gatech.edu}

\begin{abstract}
As governments increasingly adopt digital tools, public service chatbots have emerged as a growing communication channel. This paper explores the design considerations and engagement opportunities of public service chatbots, using a 311 chatbot from a metropolitan city as a case study. Our qualitative study consisted of official survey data and 16 interviews examining stakeholder experiences and design preferences for the chatbot. We found two key areas of concern regarding these public chatbots: individual-level and community-level. At the individual level, citizens experience three key challenges: interpretation, transparency, and social contextualization. Moreover, the current chatbot design prioritizes the efficient completion of individual tasks but neglects the broader community perspective. It overlooks how individuals interact and discuss problems collectively within their communities. To address these concerns, we offer design opportunities for creating more intelligent, transparent, community-oriented chatbots that better engage individuals and their communities.
\end{abstract}

\keywords{public service chatbot, civic technology, community engagement, large language models, participatory design}

\maketitle

\section{Introduction}
Public service delivery often involves mutual interaction between the government and citizens, where both sides work collaboratively \cite{makasi2021typology, lee2021crowdsourcing, osborne2020public, mehr2017artificial}. These services are closely tied to citizens' everyday lives, and their quality strongly impacts citizens' trust in their government. The most common type of service delivery consists of responding to citizens' requests \cite{whitaker1980coproduction}. For example, when a pothole needs to be fixed, a citizen reports it, and then in response, the appropriate city agencies address it. The communication channels utilized to request and provide service play an important role in shaping citizens’ perceptions of government responsiveness \cite{whitaker1980coproduction}. Governments typically utilize both traditional channels, such as face-to-face and telephone, and technology-based channels, such as mobile apps and web interfaces \cite{nl2022back}. In recent years, some cities have started to deploy chatbots, powered by artificial intelligence, as a new kind of channel \cite{makasi2021typology, abbas2022chatbots, 311chatbot2023}. While prior research suggests mobile apps and websites embody democratic ideals by increasing citizens' access to and participation in government activities \cite{levine2018citizenship, percy1984citizen, o2017uncharted}, more research is needed to explore the civic impact of public service chatbots.

Chatbots have been deployed across various fields, such as open domain \cite{medhi2017you, portela2017new, shum2018eliza, jain2018evaluating}, health care \cite{fitzpatrick2017delivering, ho2018psychological}, education \cite{sugisaki2020usability, ayedoun2019adding, coniam2008evaluating}, and customer service \cite{liao2016can, araujo2018living}. However, there is no set of one-size-fits-all design guidelines that apply to chatbots across different domains, as key design factors vary depending on each specific use case. Chatbot design considerations primarily fit into two categories \cite{liao2018all}: task performance, such as productivity and efficiency \cite{liao2016can, sugisaki2020usability, gnewuch2017towards, ashktorab2019resilient, li2020multi}, and social concerns, such as personality and proactivity \cite{fitzpatrick2017delivering, ho2018psychological}. For public service chatbots, there is limited research on how much users prioritize task-oriented versus social-oriented functions, and which design factors from these  categories matter most. 

This paper explores the experiences with and design opportunities for public service chatbots. We address the following research questions: 

\begin{enumerate}
    \item[{\bf RQ1:}] What are citizens' experiences and challenges when using public service chatbots? 
    \item[\bf{RQ2:}] To what extent can public service chatbots foster civic engagement?
\end{enumerate}
\vspace{-0.02cm}

Public service chatbots lie at the intersection of digital civics and chatbots. The scope of public services discussed in this paper refers to civic issues related to public spaces (e.g., street potholes, fire hydrant repairs, and illegal dumping). The 311 system is one of the most commonly used platforms for these services, implemented by nearly 400 cities and towns across the U.S. \cite{o2017uncharted}. We use a 311 chatbot deployed in Atlanta, a metropolitan city \cite{311chatbot2023} as a case study because its functionality is representative of the broader space of public service chatbots. We interviewed 16 participants (government officials, community leaders, and residents) and obtained the official chatbot survey data collected by the city between 2023 and 2024. From our analysis of these data, we identified three key challenges: interpretation, transparency, and social contextualization. Moreover, there is a gap between the 311 chatbot's individual-oriented design and the participants’ community-oriented practices and goals. We unpack three design opportunities: inform, report, and act collectively at the community level. Instead of enhancing chatbots’ interpersonal warmth through empathy or humor, we advocate for designs that surface real-world civic networks and ground chatbot interactions in community dynamics—as a more effective means of fostering social contextualization and civic engagement.

This paper makes three key contributions: (1) an analysis of citizens' needs and expectations of public service chatbots; (2) a set of public service chatbot design considerations based on our analysis; and (3) a discussion on potential directions for more civic engagement through public service chatbots. We hope that the challenges and opportunities highlighted in this paper will contribute to the development of more intelligent, transparent, and community-engaging public service chatbots in the future.

\section{Background}
\subsection{Digital Civics}
Digital civics aims to deploy technologies to inspire collective creativity, challenge injustice, and increase a sense of inclusion \cite{pena2017design, olivier2015digital}. This concept has been implemented in multiple civic activities, such as Internet of Things devices for better involvement in community meetings \cite{mahyar2018communitycrit}, online crowdsourcing maps for civic engagement around air pollution \cite{hsu2020smell}, and interactive installations for community connections \cite{vilaza2018here}. However, the adoption of newer technologies does not always lead to a more meaningful realization of public values \cite{van2019new, savoldelli2014understanding}. \citet{savoldelli2014understanding} identified institutional and political barriers that can impede the effectiveness of civic technology. Some scholars state that traditional interpersonal interaction is still indispensable in civic engagement, as face-to-face interaction cultivates relational and affective support, which encourages broader civic participation \cite{corbett2018problem, korn2015creating}. These insights remind us to approach digital civics with a critical perspective.

When it comes to providing public services through digital technologies, prior research has predominantly focused on e-government websites and apps. For example, \citet{silcock2001government} noted that these ``service-in-an-instant options'' free residents from the constraint of time and space. They also empower citizens through self-service administration and improve public information accessibility by removing redundant public service layers \cite{anshari2023enhancing, savoldelli2014understanding}. On the government side, such web- and app-based services reduce staff workload and increase efficiency in case handling \cite{makasi2021typology}. However, as an emerging technology, public service chatbots are rarely studied in terms of civic impact within the HCI field. Our work adopts a critical perspective to explore the impact of chatbots on civic relationships between the government and citizens.

\subsection{Chatbots}
The general applications of chatbots have been widely studied, with several researchers proposing broad design considerations and guidelines. For instance, \citet{silva2024towards} identify three principles -- naturalness, emotionality, and transparency -- based on the impact of linguistic, visual, and interactive elements on user experience. \citet{chaves2021should} investigate chatbots' social characteristics such as conversational intelligence and personification, while \citet{sugisaki2020usability} present usability heuristics for chatbot interfaces. Additional studies explore design challenges for future iterations \cite{almutairi2023chatbot} or concerns in communication and social science contexts \cite{cowan2023introduction, zheng2022ux}. Although this prior work provides useful high-level recommendations, it does not fully address the specific requirements or considerations for designing chatbots in particular applications. Chatbots for specific use cases require detailed design considerations. Prior research has provided insights in certain areas, such as balancing social and task-oriented functions in human resource chatbots \cite{liao2016can}, exploring the positive impacts of human-like cues, like language style and naming, in customer service chatbots \cite{araujo2018living}, and addressing privacy concerns related to previous utterance references in healthcare chatbots \cite{cox2023comparing}. However, limited research focuses on design considerations for public service chatbots.

In digital civics, there are two broad types of chatbots: (1) social-oriented chatbots, which enhance community interaction and facilitate civic engagement, such as the DBpedia Chatbot for answering community questions \cite{athreya2018enhancing} and the CivicBots for fostering youth societal participation \cite{vaananen2020civicbots}, and (2) task-oriented chatbots, which focus on addressing civic issues \cite{makasi2021typology}. Among various types of civic chatbots, this paper focuse on public service chatbots. These chatbots are designed specifically for facilitating government service delivery and fall into this second category. Public service chatbots have been employed in many cities worldwide, such as UNA in Latvia for enterprise registration, Kamu in Finland for immigration services, and MANDI in Queensland for neighborhood disputes \cite{makasi2021typology, abbas2022chatbots}. Prior research on public service chatbots has primarily focused on technological progress \cite{valverde2019chatbot, androutsopoulou2019transforming} and political values \cite{van2019new, makasi2020chatbot}. The development of design guidelines for these chatbots remains limited, except for work by \citet{abbas2022chatbots}, which investigates the user acceptance of the municipal chatbot Kommune-Kari in Norway. We go a step further to identify design factors that are central to meeting citizens' needs. To further explore design opportunities, we investigate whether functions from social-oriented chatbots can be adapted for task-oriented public service chatbots to enhance civic engagement.

\section{Methods}
To understand users’ expectations for and the civic impact of the 311 chatbot, we analyzed both open survey data (requested from the city's official 311 chatbot website) and interviews with 16 participants. This study was reviewed and approved by the Georgia Tech Institutional Review Board (IRB).

\subsection{Study Context}
The 311 system, introduced in Baltimore in 2001, was developed to handle non-emergency requests, such as graffiti removal or broken traffic light replacement \cite{schwester2009examination}. Currently, nearly 400 cities and towns have their own version, aiming to increase the public's trust in government \cite{nl2022back}. The 311 system uses Customer Relationship Management (CRM) technologies to efficiently and correctly collect information for processing within complex government structures \cite{chatfield2018customer}. Initially limited to traditional communication channels like phone and mail, 311 systems now support e-delivery channels, such as mobile applications, websites, social media, and chatbots \cite{nl2022back}. 

\begin{figure}[t!]
  \centering
  \includegraphics[width=1\linewidth]{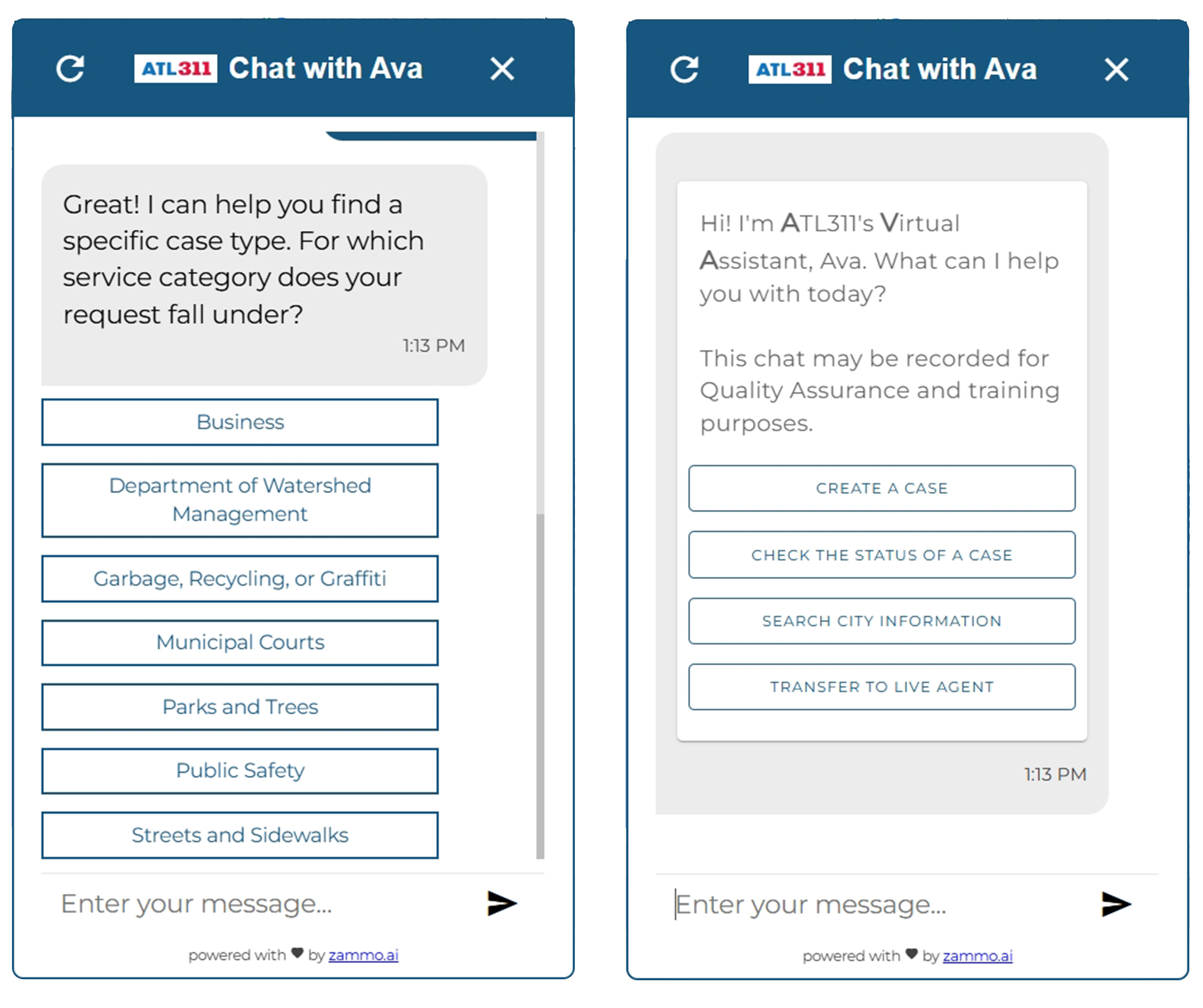}
  \caption{The 311 chatbot interface used by citizens.}
  \Description{Two screenshots of the ATL311 chatbot interface. The left screen shows a message asking users to select a service category, such as business, public safety, or streets and sidewalks. The right screen displays a welcome message from the virtual assistant with options to create a case, check case status, search city information, or transfer to a live agent.}
  \label{fig:311-chatbot-interface}
\end{figure}

The Atlanta 311 system examined in our case study was launched in 2014 and deployed its chatbot communication channel in 2023 (see Fig \ref{fig:311-chatbot-interface}). From 2023 to 2024, monthly cases reported through the chatbot channel increased from 836 to 1069 while the percentage of 311 requests processed by the chatbot (vs. other 311 communication channels) increased from 1.79\% to 2.58\%, showing it remains a minor communication channel. The chatbot is still under active development, with plans for additional functionality to be added \cite{311chatbot2023, 311chatbot2024}. This 311 chatbot serves three main functions: (1) searching city information, (2) creating nonemergency cases, and (3) checking the status of cases.

\subsection{Open Data from the 311 System}
We obtained the city's 311 chatbot survey data through a formal Georgia Open Records Request.\footnote{Under Georgia’s Open Records Act (O.C.G.A. § 50-18-70), data from the 311 system are considered public records.} The data contains 144 comments collected from January 2023 to April 2024 through a survey provided to users after each chatbot interaction. It includes both a user satisfaction score and an open-ended feedback response. The open data has the advantage of offering a high-level and comprehensive view of citizens' experiences with the chatbot. However, given the format of the survey, the collected data are limited in their ability to capture the deeper reasons behind user sentiments. To overcome this limitation, we supplemented these data with interpersonal interviews, which provide a more in-depth and contextualized view.

\subsection{Interview}
\subsubsection{Protocol}
The interviews were conducted in a hybrid format; some were conducted online via video call, and the rest were conducted in person. Each study lasted approximately 50-60 minutes. We obtained consent to record audio and video (including screen recordings) during the study. The interviews had four main parts:
\begin{enumerate}
    \item We conducted a task-based contextual inquiry session, where we asked participants to try the three major functions of the chatbot---searching city information, creating a case, and checking the current case status---while thinking aloud \cite{van1994think}.
    \item We conducted a semi-structured interview, where we asked users about their overall experience with the chatbot, their satisfaction with the responses from the chatbot, and their perception of the chatbot compared to other 311 channels and general search engines.
    \item We conducted a participatory design section \cite{muller1993participatory}, encouraging participants to explore the question, `What do you expect a chatbot to be like?' To facilitate the discussion, we first introduced five conceptual design scenarios and asked participants' opinions on potential features, such as critical community incident reminders (e.g., road closures, water shortages) and public event engagement advertisements (e.g., local elections, community facility redevelopment, and sustainability-related public lectures). Participants then shared their current community practices and how they envisioned the chatbot might facilitate civic engagement.
    \item We asked users about how their experience with the 311 chatbot affects their opinions about civic relationships.
\end{enumerate}
The first two sessions aimed to explore key design factors of the chatbot (RQ1), while the latter two focused on assessing its current civic impact and identifying opportunities to enhance civic engagement (RQ2). We also conducted a 30-minute follow-up interview with four community leaders involved in the first round of the study to test the hypotheses derived from previous interviews.

\subsubsection{Participants}
To ensure participants came from geographically relevant locations, we recruited individuals through two channels: location-based online platforms (Nextdoor and Facebook community groups), and the city's  Neighborhood Planning Units (NPU) meetings, which involve citizens in neighborhood municipal discussion and management. We conducted a survey to gather their demographic information (gender, age, race, years living in the city), and their experience with the 311 system(frequency of use and preferred communication channels). The participant selection criteria ensured representation across diverse demographic and civic engagement characteristics, as summarized in Table \ref{tab:participant_information}. Participants ranged from less engaged individuals---those who self-reported never using the 311 system and rarely attending community meetings---to highly engaged residents, community organizers, and government officials.

\begin{table*}[h]
\centering
{\small
\begin{tabularx}{\textwidth}{X X}
\hline
\textbf{Demographic Information} & \textbf{Participant Counts} \\
\hline
Self-Reported Racial Identity & White (9), Black or African American (4), Asian (3) \\
Age & 20-29 (4), 30-39 (2), 40-49 (2), 50-59 (4), >60 (4) \\
Gender & Male (7), Female (9) \\
Length of Residency in This Metro City & More than five years (14), One to five years (2) \\
311 System Usage Count & >5 times (11), 1-5 times (3), Never (2) \\
Community Role & Government Official (1), Community Leader (4), Resident (11) \\

\hline
\end{tabularx}
\caption{Participant Information}
\Description{A table summarizing the demographic information of 16 participants. It includes self-reported racial identity (White, Black or African American, Asian), age groups (ranging from 20–29 to over 60), gender (male and female), residency length in the metro city, 311 system usage count, and community roles such as government official, community leader, and resident.}
\label{tab:participant_information}
}
\end{table*}

\subsection{Analysis}
For the open data, we categorized the 144 comments on chatbots into 11 themes, such as location input errors, follow-up needs, and tedious interaction dialogues. For interview data, we held interpretation sessions \cite{beyer1999contextual} to collaboratively synthesize findings from our contextual inquiries and interview notes. After completing all 16 interviews, we conducted a thematic analysis \cite{inbook, beyer1999contextual} of approximately 12 hours of video recordings. To ensure the reliability of the qualitative analysis, two researchers independently coded each participant's data \cite{mcdonald2019reliability}. Each researcher generated around 50 open codes for one participant's transcripts. In the early stages of the codes-to-themes process, we generated 62 preliminary themes and categorized them into five domains: participants' needs, feedback on the interaction design, differences between the 311 chatbot and other channels,  design opportunities on proactive functions, and reflections on civic relationships. In parallel, two authors integrated the findings from the 144 open data comments with the synthesized interview findings. After further review and discussion of all observations and themes, we identified three design features of the 311 chatbot and its impact on civic relationships, which we discuss in the next section.

\subsection{Limitation}
One limitations of our work is that most of the participants demonstrated interest in civic engagement, even the two who reported never using the 311 system. This may have biased the findings toward perspectives that are more favorable to expanding the role to the 311 chatbot in fostering civic participation. Citizens with lower interest or involvement in civic issues may hold different expectations regarding the purpose and functionality of such systems. Future work could explore these perspectives through qualitative studies focused on less engaged populations, examining whether and how social-oriented chatbot features might spark greater interest in civic matters.

\section{Findings}
Through our analysis of the interview and survey data, we identified three key challenges for the current 311 chatbot: interpretation, transparency, and social contextualization. In addition, its current design emphasizes individual task efficiency but lacks important community-oriented capabilities. We propose three design opportunities for the chatbot: inform, report, and act to raise its social awareness and enable it to better support community engagement. The individual-level challenges and community-level opportunities are summarized in Table \ref{tab: summary}. In the following discussion, we denote interview quotes using ``P'' and survey comments using ``S''.

\begin{table*}[h]
\centering
\caption{Summary of Key Challenges and Considerations for the 311 Chatbot}
\Description{A table outlining challenges and corresponding design considerations for the 311 chatbot at individual and community levels. Individual-level challenges include interpretation, transparency, and social contextualization. Community-level challenges are categorized as inform, report, and act. Each challenge is matched with design opportunities, such as improving language understanding, enhancing transparency in request processing, supporting collective reporting, and enabling community-driven actions.}
\label{tab: summary}
\begin{tabular}{|p{1.5cm}|p{6.8cm}|p{7cm}|}
\hline
\textbf{Level} & \textbf{Challenges} & \textbf{Design Opportunities and Considerations} \\
\hline
\multirow{3}{*}{Individual} 
& \textbf{Interpretation}: Unable to understand complex requests; constrained input flow; vague or outdated responses; lack jurisdictional awareness of locations
& Enhance interpretation performance to deal with detailed and complex requests, and provide precise and up-to-date outputs in a natural conversation flow \\
\cline{2-3}
& \textbf{Transparency}: Opaque request processing compared to manual channels; limited case updates and prioritization clarity 
& Tackle broader systemic problems in the 311 bureaucracy to provide updates, including assigned departments, timelines, and prioritization logic \\
\cline{2-3}
& \textbf{Social Contextualization}: Lack of empathy and conversational flexibility; perceived as a transactional tool, weakening social connection
& Balancing efficiency with social engagement, integrating social-oriented features \\
\hline
\multirow{3}{*}{Community} 
& \textbf{Inform}: Reactive information provision; citizens reluctant to seek local or social information perceived as irrelevant, hindering community engagement
& Support proactive, customized (i.e., location-aware and case-driven) information dissemination while respecting privacy and avoiding harmful bias\\
\cline{2-3}
& \textbf{Report}: Lacks support for existing community practices of collective reporting, a more effective driver of social connection than artificial empathy
& Aggregate and visualize community-wide issues, and support collective escalation, while managing fraudulent reports and respecting diversity \\
\cline{2-3}
& \textbf{Act}: Fails to facilitate community-driven actions beyond formal case resolution; struggles to mobilize broader public participation
& Suggest relevant activities at reporting time; enable community leaders to gather interests and plan initiatives from diverse groups through citizen reports \\
\hline
\end{tabular}
\end{table*}

\subsection{Individual Level}
Based on our survey and interview data, citizens individually experience three key challenges when using civic chatbots: (1) issues where the chatbot does not make a correct interpretation of their request, (2) lack of user transparency into how and when their requests are handled, and (3) the inability for chatbots to leverage social contextualization that exacerbates detachment between citizens and the government.

\subsubsection{Interpretation}
We examined the 311 chatbot's interpretation performance, which refers to its ability to understand and appropriately respond to citizens' requests. Public service chatbots must deliver high interpretation performance to handle the complexity of user requests, ensure accuracy, and provide reliable, up-to-date, and comprehensive outputs when addressing civic issues.

The input process for the 311 chatbot is tedious and unnatural due to its low interpretation performance, which falls significantly short compared to human representatives. When submitting requests, citizens must provide sufficient information for the government to fully understand the situation, often requiring multiple details. The chatbot is tasked with extracting key information from these details and categorizing the request precisely. However, the current system achieves this in an inefficient manner: after describing their issue, citizens are required to confirm the chatbot’s interpretation by selecting from a lengthy list of options that best match their request, often through multiple rounds of dialogue. Users have to restart the process and repeat the tedious steps if a mistake occurs. As S91 remarked, \textit{``Why do I have to play 50 questions to get my garbage collected?''} Moreover, the chatbot forces users to frame their requests in a restricted and unnatural way. For example, as S16 noted in the survey, \textit{``My problem was primarily with the sidewalk, but I had to say it was the street in order to have a case.''} The current system relies on a keyword-matching algorithm that only recognizes predefined terms in its database, requiring users to guess the correct keywords. At the start of a conversation, however, citizens often frame complex requests with uncertainty and ambiguity, making it difficult to select the right words at this stage, which violates the principle of natural interaction \cite{ruoff2023onyx, horvitz1999principles, maclellan2018framework}.

As a result of low interpretation performance, the chatbot's task completion outcomes heavily depend on how users phrase their requests. In contrast, human representatives can effortlessly understand situations without requiring lengthy dialogues and can tolerate ambiguity based on contextual cues and their prior experience. Citizens perceive this disparity as a technological gap rather than a failure of government operations. Yet they expect the chatbot to perform at least as well as a human representative---or better. For instance, one participant envisioned the chatbot capable of automatically creating a case by analyzing the content of an uploaded photo, eliminating the need for additional user input.

The output of public service chatbots requires precision, which means providing the exact information needed by citizens, including details such as the address, time, and issue category, which the current chatbot is unable to provide. If there is a lack of exact data, instead of saying ``I don't know'', the chatbot responds with \textit{``either not even remotely answering or somewhat close, but still not the question that you need to have''} (P8). For instance, when P10 searched for information on the process of getting a city business license---which had changed in recent years---the chatbot responded with a link containing general information about business licenses rather than directing them to the updated procedure and required forms. This response was described as ``not helpful.'' Given that policies and regulations can change over time \cite{chatfield2018customer}, chatbot responses should accurately reflect these changes. As a result, P10 changed back to the traditional channels, calling 311 to get the answer. Unfortunately, having to complete a task across multiple channels---to ``double step''---increases the time users spend on the task, which degrades the user experience and decreases trust in the chatbot. \textit{``Speaking to [a live agent] through the chatbot when it starts going wonky''} (P15), can be an effective backup when the chatbot fails, letting citizens resolve their issues entirely within the chatbot channel.

Territory is a significant component of public service chatbots \cite{o2017uncharted, o2014caring, o2012managing}, as it is closely tied to the jurisdiction of specific local governments. However, the chatbot struggles to interpret the complexities of administrative divisions within civic systems. In 8 survey comments and 10 interviews, citizens failed to report because of location input. Specifically, when the chatbot asked ``In which city?'' all of the interview participants felt confused and entered ``Atlanta'' in response. However, users assumed that ``Atlanta'' referred to the larger Metro area, but the ``city'' in the question actually refers to the municipality (the ``smaller city''). The Atlanta Metro area has 140 municipalities \cite{Atlanta}. The Metro area is used in most daily life scenarios. However, municipalities—though less commonly recognized by residents—serve as the basis for government jurisdiction and the division of responsibilities and powers. `City A' cannot provide a chatbot service for `City B,' even though both are part of the Atlanta metropolitan area. Accurately interpreting the responsible government from geolocation inputs and effectively explaining complex administrative divisions remain significant challenges for the 311 chatbot.

\subsubsection{Transparency}
In the context of the 311 chatbot, transparency means revealing the bureaucratic mechanisms behind the handling of user requests (e.g., if I submit a case to fix a public transit bus stop, how will that case be handled and prioritized, and by which departments?). Reported case information that is accessible from officials via traditional channels is currently inaccessible through the chatbot. For instance, P7 shared the experience that the chatbot was unable to find the case using the provided number, but the 311 representative on the phone successfully found it and explained the case status. If citizens go to NPU meetings, they can learn the whole procedures of case resolution, as P6 mentioned, \textit{``[How 311 deals with a case] was more so that the public didn't understand the process. I understood only because I come to NPU meetings.''}

When using the case status check function, the chatbot only shows whether the case is closed or open. Citizens never receive follow-up information from the 311 chatbot, as P4 complained, \textit{``I have absolutely zero faith that I will get a response. It said the due date [of my case] was two weeks from the day you put it in, but you never get an update. It is like going into this black hole and [the case] will disappear as 311 closes it.''} To facilitate follow-ups, the chatbot should have access to the broader 311 system's databases, enabling it to link the current case number with the corresponding work order in the relevant department. From the interviews, citizens identified three key pieces of information they expected when using the status-checking function: the assigned department, the current work order status, and the anticipated timeline.

When shifting case reporting from human-to-human interaction to a fully automated process like a chatbot integrated with the CRM system, it is unclear whether and how cases should be automatically prioritized and escalated. Currently, 311 officials make these decisions manually. For example, P3 shared an experience at a community meeting where a damaged sidewalk section was reported. The city council member for the district deemed it a priority and decided to address it when funding became available. In contrast, cases reported through the 311 chatbot lack explicit prioritization rules. As P11 noted, \textit{``That's an excuse for not having services in the street because they'll only send people to where there is enough data about potholes. What's their threshold? I don't know. Maybe five people have to complain about that in order to get it to be a valid data point.''} The fully automated system lacks the analytical capabilities for prioritizing cases that are repeatedly raised by multiple residents or that remain unresolved for a long time.

According to P16, a government official, the current CRM automatically closes cases reported by the chatbot due to predefined deadlines even if the case has not been addressed. This automatic closure without explanation raises significant concerns about equality and transparency, highlighted across multiple survey comments and interviews. To address this, clear regulations on prioritization must be established, and an escalation method should be embedded in the chatbot or in the case management system. Moreover, as S67 stated, \textit{`If overdue, then additional reporting/action should be indicated: ``We recognize that this item is overdue. We will ask that a customer agent reach out.'''} Incorporating human oversight is essential to ensure fairness, transparency, and accountability in such an automated system. Furthermore, the entire 311 bureaucratic system functions as an `information desert' and remains opaque \cite{lee2020toward}. Addressing the transparency issues of the chatbot, therefore, requires tackling the broader systemic problems.

\subsubsection{Social Contextualization}
Compared to human representatives, the chatbot lacks social contextualization, which refers to the chatbot's ability to produce social behaviors, such as empathy, humor, or small talk, and build connections with users \cite{chaves2021should, bjorkqvist2000social}. The chatbot cannot show empathy, and it treats people as data points. \textit{``It's only data. There's no human component. Submit a case about scooters littering the sidewalk, so I've got to step over them. The data doesn't. The AI doesn't care no matter what I feel. But if I send a picture and a human being sees it, they say, `Oh my gosh, look at what that woman has to deal with!'''} (P11). When experiencing bad civic issues, people want to receive empathy from others. Additionally, citizens tend to talk more and get more detailed information when interacting with another human than when interacting with a chatbot. As P8 said, \textit{``If you're asking the question, you're realizing, Oh no, I need this too. Thank you for giving me this information, but what about this? What about it then? It becomes more convoluted and complicated as you go on because you might think `I just need this one question answered.' But then, when it’s answered, you realize you have additional things you might need answers as a result.''} A convoluted conversation with the 311 representative allowed P8 to gain more knowledge about the city's business license regulations. The communication flow in interpersonal interaction is unconstrained and flexible. It enables citizens to iteratively negotiate on a case or gather information by asking questions they did not initially consider. In contrast, the chatbot's conversation flow is unilateral and constrained. The response to each request is predefined, limiting further exploration. 

This lack of social contextualization leads citizens to perceive the 311 chatbot as a tool instead of a civic communication channel. Participants want to differentiate the 311 chatbot from humans. P13 indicated, \textit{``I appreciate that it's a computer. I like to keep the difference. It doesn't have to be more human-like in that case I would just call 311.''} Citizens are reluctant to spend extra time interacting with it. Users often leave the chatbot interface once their individual purpose is fulfilled, ignoring additional prompts. Thus, the chatbot is reactive and circumstantial: citizens only use it when they need to address specific civic issues. \textit{``Because what you really want to do is to help improve the efficiency of the city. So the total benefit to the city is if I can do the work of that individual, then they can process more cases faster at a cost less to the city''} (P12). P10 noted, \textit{``I don't think it increased civic engagement, because, at least as it exists now, [the 311 chatbot's purpose] is to solve problems.''} Efficiency dominates other design factors, making the 311 chatbot purely task-oriented and devoid of social-oriented features.

The current efficiency-driven design might be decreasing users' social connectedness. Compared to the chatbot-based interaction, human-to-human interaction transforms public service from transactional to relational, fostering connections between citizens and government officers or community organizers through each conversation \cite{osborne2020public, o2017uncharted, lee2021crowdsourcing, hartmann2017citizen}. P1 stated: \textit{``Growing up in a small town, I would go to the town hall. I would pay my taxes with a paper check. I would talk to the town clerk. I would know everybody. Make a point to know everybody in the town hall. Make human connections with them. And I feel that I'm not there yet with the Chatbot. Maybe I can get there. But I still feel like it's another barrier to getting things done.''} In contrast, the current chatbot design isolates individuals and severs connections with government representatives and other citizens. As stated by P11: \textit{``We are isolating the people and they're just becoming bots themselves. We are forcing them to give up the human component of [public service] jobs and to be looking only at data and to be measuring [government] performance only by the data.''} There is a contradiction over street-level bureaucracy: on the one hand, service is delivered by people to people, invoking a model of human interaction, caring, and responsibility. On the other hand, service is delivered through a bureaucracy, invoking a model of detachment and equal treatment because of resource limitations and constraints \cite{lipsky2010street}. Transferring from human-human interaction to human-chatbot interaction exacerbates this contradiction: simplifying and routinizing public service, and aggravating detachment among individuals. It forces citizens to focus too narrowly on their own issues, missing the chance to learn about their community and local government.  While the chatbot aims to improve accessibility to public services and address the long hold times of hotline services, efficiency does not necessarily mean sacrificing engagement opportunities. The next section will unpack the redesign opportunities of public service chatbots to strengthen community bonds and enhance civic engagement.

\subsection{Community Level}
In this section, we focus on the community-level design challenges and opportunities relevant to public service chatbots. At the community level, we refer to collective sensemaking activities by groups using these chatbots to seek information or report on concerns relevant to their local neighborhoods. Currently, there is a gap between the 311 chatbot's individual-oriented design and the participants’ community-oriented practices, goals, and approaches. We find that people often report on community-facing issues (e.g., street potholes, fire hydrant repairs, and illegal dumping) instead of personal issues related to their property (e.g., driveway pavement, broken water pipes, lost trash cans). For instance, P1 said, \textit{``This would be a watershed [i.e., possible water leak and main break of the water system] and I want to report, not my own house. But I want to report that I see a leak on a city street.'' } The sense of community is integral to the 311 system’s functionality. Indeed, the originally described intent of the 311 system was cooperative: \textit{``311 is a coproduction program in that it introduces a collaborative model for the maintenance of the urban commons or the public spaces and infrastructure of a neighborhood. Commons are everyone’s problem but no one’s formal responsibility''} \cite{o2017uncharted}. In practice, the community already collaborates to identify and report issues, which we will further demonstrate in the following sections. 

The design of the 311 chatbot should reflect its original community intention and the community members’ collaboration practices. If designed in that way, the 311 chatbot could offer potential solutions for community engagement challenges. Public service chatbots have the potential to increase community connections and engagement, as they tackle practical civic issues relevant to all community members in their daily lives. Based on our interviews, we identified three design opportunities to move towards a more community-oriented approach: (1) inform, redesigning the 311 chatbot’s information search function to increase local social awareness; (2) report, enabling community members to report and escalate issues collectively; and (3) act, organizing social activities to address concerns or informing policy based on the commonly reported issues.

\subsubsection{Inform}
There is a redesign opportunity to shift from the current reactive searching (i.e., when users are only given content they explicitly requested) to proactive disseminating (i.e., when systems provide additional non-requested, but potentially useful, context) to encourage social awareness. The information obtained through reactive searching, as implemented by the current chatbot, strictly follows the user's request and does not offer additional, potentially useful context. Since citizens are often reluctant to take the initiative to learn information that seems distant or peripheral to their routine activities, reactive searching might hinder users from accessing information outside their usual awareness but can be very valuable in their daily lives. For instance, P8 did not know that local kiosks in his community could help residents search for jobs and housing events, even though they had been deployed for years. Awareness of certain information is often a prerequisite for citizens to engage in public affairs, especially for community collaboration. P12 expressed common challenges shared by residents, \textit{``We gotta understand that most citizens don't understand city government and city government can be very complex. Most citizens who are not engaged, just have basic questions. In my case, they ask about what’s the right meeting to attend, whether it be a City Council meeting or an NPU meeting; and who is my Council member and how to contact them.''} Therefore, it is essential to proactively anticipate and offer unsolicited information as a way to disseminate community information, enabling citizens to respond more effectively to potential engagement opportunities.

Customization is a key factor of proactive function design. Proactive community notices offered by current platforms (e.g. from the local newspaper's app and Nextdoor) are drowned out by irrelevant information due to the lack of customization. Previous research discusses multiple customization approaches, such as adapting chatbot functionality, interfaces, and contents to users' preferences \cite{neururer2018perceptions, medhi2017you, jain2018evaluating, fan2006personalization, hwang2019data}. Based on the prior literature on chatbot proactivity \cite{tennenhouse2000proactive, salovaara2004six, morrissey2013realness, jain2018evaluating, duijst2017can, brandtzaeg2017people}, we identified a specific design opportunity for public service chatbots: customized proactivity that is location-based, case-driven, and context-aware.

People approach the 311 chatbot with specific questions or unresolved issues that reflect their interests and serve as a reference for customization. For example, P9 suggested embedding specific information directly in the current dialog. \textit{``Maybe you're submitting a new business permit or holding an event down there. It has a specific prompt: sending your suggestions on the south downtown design and activation plan and the government. It would have to be more targeted, as opposed to just general (requests)''}. The chatbot could conclude with community news related to the previous dialogue. This customized information, provided naturally within the conversation, enhances the flow of the dialogue and strengthens the social contextualization discussed in Section 5.1.3. Community informing helps citizens gain a deeper understanding of their community, fostering social awareness.

There are several challenges to overcome when thinking about designing to inform the public. Customization based on conversation history raises privacy concerns \cite{cox2023comparing}. Privacy regulations should be established, ensuring that sensitive information is not shared publicly and that the types of customizations implemented are carefully evaluated for appropriateness. In addition, harmful bias could influence customizations. For instance, it might reflect poorly if the system provides information about jail hours in one community while offering guidance on processing a new business license in another. Lastly, designing proactive functions, particularly within the conversation, should not disrupt users' primary tasks or overwhelm them with excessive information.

\subsubsection{Report}
At the community level, report refers to working together to identify, document, follow up, and escalate issues. While the current design does not reflect community connections, the community already works together to identify and report issues. Displaying these connections when reporting a case creates a sense of being heard by others and motivates greater engagement. From the interviews, we found evidence of existing collaborative practices, including discussing civic issues, reporting them collectively, and assisting others in submitting requests. \textit{``A lot of times I have neighbors that will text me to do it for them. They don't use technology as much that's definitely a hindrance''} (P15). \textit{``There was a fire hydrant that I reported leaking back in August of this year. Others also reported the fire hydrant, doing the same thing''} (P6). The current chatbot does not yet support these community collaboration practices, making citizens feel like they are \textit{``yelling into an outer space and no one hears it''} (P8). One design recommendation is to show the user the number of other people who report the same issue. This method would encourage social contextualization in a fashion that is quite different from prior research, which emphasizes having chatbots display human-likeness cues, such as temperament, sense of humor, or small talk \cite{portela2017new, meany2010humour, brandtzaeg2017people}.

Unlike simulated empathy provided by a chatbot, this method connects users with real neighbors, fostering a sense of being heard by others and strengthening community bonds. Moreover, these bonds are key motivations for civic engagement. P3 said, \textit{``It gives [people in the community] an idea of the scale of the impact. If you’ve already reported it, you’re probably not going to do anything else, like reaching out directly to the city councilperson or the specific department, unless you have a compelling reason to, such as seeing that the issue affects many others. It could give you the push to take further action.''} Showing community connections reflects the essence of 311's co-production model \cite{o2017uncharted}, where reporting is not only for individual benefit but also serves to help others facing the same civic issue. Highlighting the broader community impact could motivate citizens to more actively identify and report issues in the future, as well as take further actions, such as follow-ups and escalations.

Public service chatbots offer an opportunity to visualize synthesized community-wide concerns and enable collective follow-up. Reporting at the individual level overwhelms government officials and community leaders with ``so many calls and personal talks.'' The only channel for them to synthesize common issues is through community meetings. During monthly NPU meetings, residents and community leaders raise long-standing unresolved issues to draw 311 officials' attention and accelerate their resolution. However, these community meetings may \textit{``suffer from a lack of data. There aren't a ton of people who come to the meeting. So, it's hard to come to any meaningful data conclusion. We don't know what other 311 reports are being filed [by those not at the meetings]''} (P3).

No one has a clear understanding of which issues are community-wide concerns. To address this, the bureaucratic system should include a feature that collects all reported cases and synthesizes community-wide issues. Within this bureaucratic system, the chatbot could aggregate and share the most frequently reported issues back to users. For example, P8 described a potential feature that visualizes aggregated information, \textit{``We could see it in real-time, like there's a red bar and it goes higher and higher based on how many people have reported the issue. Once it's completed there's something green.''} Compared to community meetings, visualizing community issues on the chatbot increases transparency and accessibility of reported data, and encourages collective follow-up actions and accomplishment. The public can track common issues, facilitating the escalation process and ultimately leading to case resolution. 

The design of community-level reporting faces several challenges. First, it should address divergences within the community. For instance, as P8 noted, \textit{``Due to gentrification, there are more people in [the community] that are two-parent families and double incomes, so they're expecting one thing, and then the other people who are working class have to deal with other things. People have told me that they actually have separated and they operate differently.''} P8 also highlighted the differing perspectives on bus delays between these two groups of community members. Community members that moved into the area during gentrification typically drive rather than use public transportation and are less likely to report bus delays and similar concerns. In contrast, working-class citizens, who rely more on public transit, advocate for significant efforts to address transit concerns. Citizens from diverse socioeconomic and cultural backgrounds will often have differing opinions on what constitutes an important community issue that should be prioritized. Second, the chatbot should consider the potential for fraudulent reports when operating at the community level. For example, an external agent might attempt to create a problem within the community, or individuals might use inflammatory language to elevate an ordinary issue to the level of a significant community concern. Third, collective reporting does not imply that a single person should report all issues on behalf of the community. The 311 system requires individual data to determine its importance, and it is also important to avoid overburdening community leaders.

\subsubsection{Act}
Taking actions from a community perspective goes far beyond just submitting cases to the local authority and extends to the extra work needed to make the community better. In the current 311 system, acting typically means the relevant city department fixing the reported issues, which is different from acting to work towards broader community goals. For instance, P4, the NPU chair, suggested a kind of community act, \textit{``Well, let's say you go to the playground and the swing is broken. If there were a lot of reports like that. And then, if somebody says, `Let's design a new playground. Do you want to be involved, you know, and then there are efforts to raise money for this? What you know you can help by doing.'''} The city's action over public service is the baseline while the community acts as an intermediate level to bridge gaps in the baseline (e.g. untimely problem-solving) or conduct extra work to enhance the community.

A key challenge in community actions is the lack of a critical mass of engaged community members or volunteers. As P6 noted, \textit{``I think engagement can be a main problem for the community. When you look at the people who volunteer, it's always the same people. People in their 60s and 70s and 80s are the backbone. But since the pandemic, they're not volunteering as much anymore. Then, younger people don’t participate because of traffic, family obligations, and work. The whole concept of volunteering isn't what it used to be.''} In light of this challenge, six participants highlighted the potential for the 311 chatbot to engage more people in community activities. 

Suggesting community activities tailored to citizens' requests could increase their likelihood of participation. This approach is similar to the customized proactive informing discussed earlier, but takes it a step further toward action, linking the dialog to a specific social activity, and collecting the user's interests. P6 described a potential dialog example, \textit{``I think it could be when someone's reporting, for instance, graffiti. Questions can be: where is it? Do you have any pictures? And then the last question is, would you be interested in helping a cleanup effort?''} Moreover, P3 suggested a calendar function to collect availability to schedule community activities in advance. This feature would provide citizens with activities at hand, eliminating the gap for individuals who report and care about that issue but don't know how to participate. It transforms the simple reporting process into a more practical and engaging experience.

Community leaders could also benefit from more act-based mechanisms in the 311 chatbot. They might use reported cases and community members' participation intentions to decide whether to organize social activities and to determine the nature of those activities. P10 said, \textit{``depending on the level of what it is. The graffiti has been here. Maybe there's a tipping point in the number. And there are 20 people. Is the community going to be the one to clean it up? But if it's 40 people, is it going to be the city to set it up?''} Furthermore, community leaders can organize activities targeted to certain groups of people based on demographic data. For instance, if community leaders identify that younger generations are less engaged, they can design activities that are more attractive to them.

\section{Discussion}
Our findings highlight the challenges and design opportunities for the current 311 chatbot from both individual and community perspectives, inspiring broader considerations for public service chatbots exemplified by the 311 chatbot. In this section, we explore the future of public service chatbots, including technical opportunities, civic potential, and implementation challenges.

\subsection{Can Large Language Models Improve Interpretation and Reasoning?}
A major technical barrier for public service chatbots is improving their interpretation performance, which is essential for both the individual-level user experience and the feasibility of the proposed community-level features. Large Language Models (LLMs) are the most promising technology to improve language processing performance \cite{vaswani2017attention, brown2020language, makasi2021typology}. However, they are not a panacea for public service chatbots.

At the individual level, response accuracy is crucial for public service chatbots, as correctly interpreted requests are essential for assigning cases to the right department, and citizens often rely on search results to make decisions. Accuracy cannot be guaranteed due to the propensity for LLMs to produce inaccurate responses (i.e., hallucinations; bullshit) \cite{xu2024hallucination, robert_2024, ji2023survey, hicks2024chatgpt, zhang2023siren}. These limitations are especially pronounced in complex or domain-specific contexts such as social services \cite{wang2023evaluating, ziems2024can}. As public service chatbots aim to guide citizens' decision-making, such misinformation risks real-world consequences and may erode trust in civic systems.

At the community level, the feasibility of case-driven and context-aware functions largely depends on the chatbot's logical reasoning capabilities to connect available information. Chatbots must determine which community information is relevant to the user's interests, otherwise irrelevant information may annoy citizens. Although multiple prompting methods \cite{wei2022chain, yao2023tree, besta2024graph} have been introduced to enhance reasoning ability, LLMs still fall short of human-level performance in reasoning \cite{huang2022towards,valmeekam2024planning}. Specialized LLM approaches, such as fine-tuning \cite{yang2024harnessing, weber2024large} or input-parsing \cite{bai2023constituency, valverde2019chatbot}, show potential, but future studies are needed to test their feasibility in the public service area.

\subsection{How Can Chatbots Support Civic Engagement?}
While prior research often frames 311 chatbots as functional, task-oriented tools for service delivery \cite{makasi2021typology}, we position them as socio-political infrastructures. The shift from human-human to human-chatbot interactions in public service channels can impact the distribution of power, meaning a citizen's ability to achieve their own goals and ends via bureaucratic systems. Public service is a street-level bureaucracy, where frontline workers interact directly with citizens and have extensive discretion in the execution of cases \cite{lipsky2010street, terry1993we, stickley2006should}. Compared to officials, citizens have limited control after case submission. Introducing chatbots in public service may exacerbate such power imbalances and further diminish the power of citizens as they lose interpretive flexibility, face opaque procedures, and lack channels for negotiation or escalation. Viewed through the lens of social translucence---digital systems that make socially relevant information visible \cite{erickson2000social}---the current design creates a vacuum of interaction, stripping away the visibility, mutual awareness, and social cues that sustain accountability in public systems. Citizens submit cases into isolated, windowless channels, unable to see whether others share their concerns, whether officials have taken action, or who holds responsibility. The result is not just a usability problem, but a breakdown in civic responsiveness and democratic transparency.

\begin{figure*}
    \centering
    \includegraphics[width=0.7\linewidth]{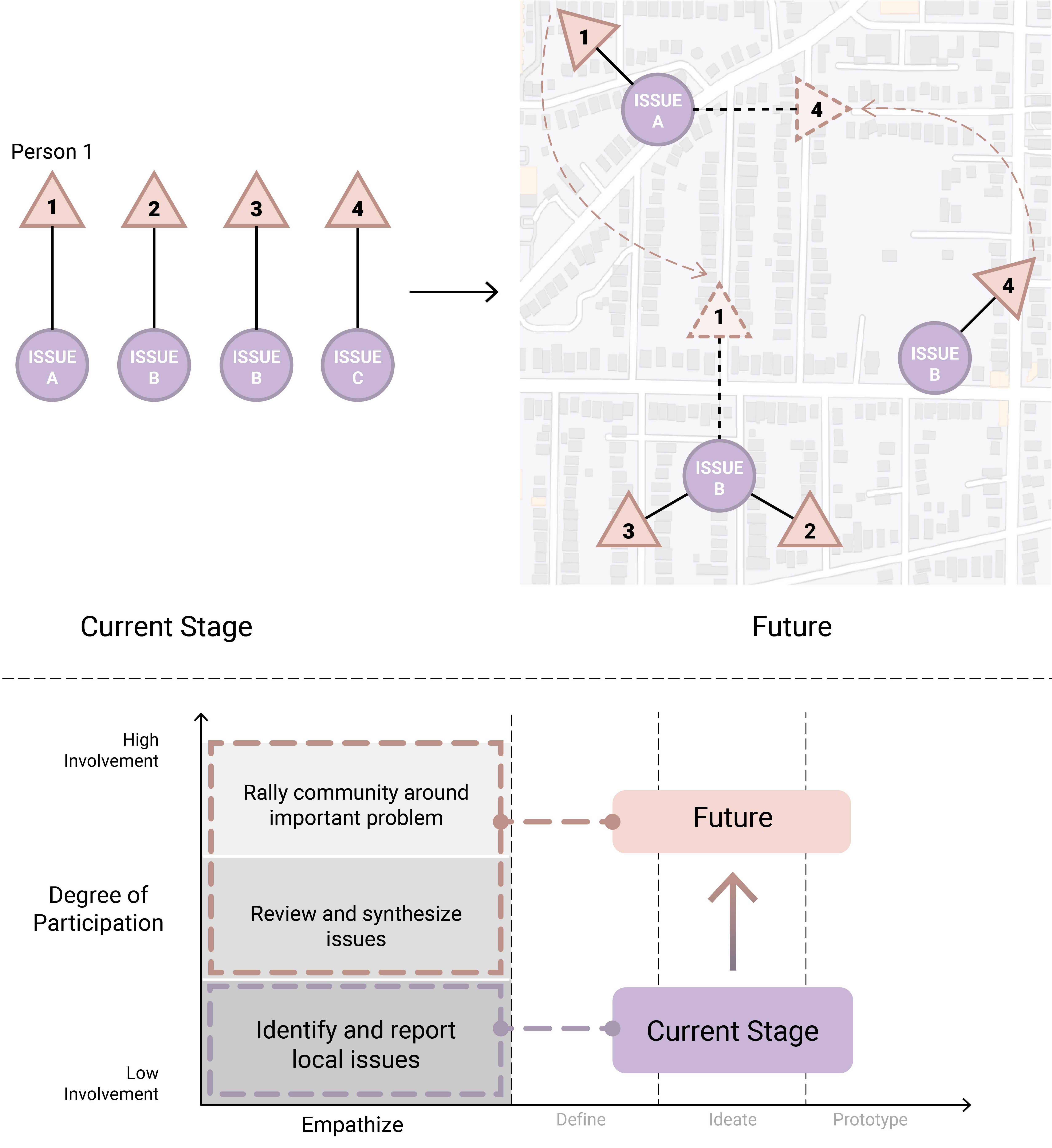}
    \caption{ Erickson and Kellogg's \cite{erickson2000social} Social Translucence framework (top)  and Reynante et al.'s \cite{reynante2021framework} Open Civic framework (bottom) along with a depiction of where future research could go within these frameworks to better support civic engagement.}
    \Description{The top section contrasts current and future chatbot designs. On the left, a single person reports multiple issues in isolation; on the right, those reports are spatially distributed and collectively connected across a map. The bottom section shows a layered chart based on the Open Civic Framework, illustrating civic participation from low to high involvement stages: empathize, define, ideate, and prototype. The current chatbot design supports only low-involvement 'identify and report' activities, while future research aims to enable higher-involvement activities like synthesizing and rallying around community issues.}
    \label{fig:framework}
\end{figure*}

Moving beyond the vacuum-like model, we argue that public service chatbots should support more complex social connection, including relationships among community members, between communities and the government, and across different civic groups. Prior research focuses on simulating human-like qualities, such as temperament, humor, or small talk \cite{portela2017new, meany2010humour, brandtzaeg2017people, ho2018psychological, ayedoun2019adding, liao2016can, liao2018all}. However, these superficial cues offer limited value in civic settings, where authenticity, coordination, and accountability are critical. Instead of mimicking interpersonal warmth, we call for designs that surface real-world civic networks.  Guided by social translucence \cite{erickson2000social}, our proposed community-aware features reconfigure the chatbot’s role as a social proxy, shifting from an individual-to-case orientation to an individual-to-community model (see Figure \ref{fig:framework} upper part).  This shift enhances visibility into shared issues, fosters awareness of community dynamics, and clarifies stakeholder accountability. Beyond service delivery, the chatbot becomes a site for community knowledge building and collective action.

Among civic participation frameworks \cite{arnstein1969ladder, fung2006varieties, whitaker1980coproduction, drobiazgiewicz2018role, reynante2021framework}, we adopt Reynante et al.'s \cite{reynante2021framework} Open Civic Design as our primary analytical lens to examine the role of community-level chatbots in the civic participation process. Open Civic Design divides the public decision-making process into four phases: empathizing, defining, ideating, and testing \cite{reynante2021framework}. Given its case-driven feature, we view the public service chatbot as a tool in the initial stage---one that identifies community problems, as shown on the lower part of Figure \ref{fig:framework}. The individual-level chatbot enables the identification and reporting of local issues, whereas community-level chatbots go beyond simple reporting by synthesizing common issues and rallying the community around key problems. This approach ensures that citizens can participate in different capacities, thereby expanding the scope of civic engagement. Furthermore, Open Civic Design conceptualizes community problem-solving as an iterative process \cite{reynante2021framework}. City fixes serve as a baseline---not as an endpoint, but as a starting point for further engagement. This aligns with our finding that community members view action not merely as case submission, but as an ongoing effort to improve their neighborhoods (Section 4.2.3). The chatbot can support this iterative process by acting as a feedback platform---gathering citizen input, tracking progress, and providing real-time insights---thereby deepening civic engagement over time.

\subsection{Are Community-Level Features Feasible?}
While our findings highlighted the design opportunities of community chatbot features, their practical implementation presents a range of technical, organizational, and ethical challenges. From a technical perspective, beyond the interpretation and reasoning limitations discussed in section 5.1, a deeper barrier lies in integrating community features into existing 311 CRM systems. While community-aware features rely on broader data access and public visibility, they face systemic constraints: the 311 infrastructure often functions as an information desert, lacking accessible or actionable data \cite{lee2020toward}, and remains fragmented across siloed departments and legacy systems \cite{nam2013success, laaudit2021}. In New Orleans’s case, such integration was described as ``extraordinarily complex,'' requiring substantial engineering effort and manual alignment with internal workflows~\cite{nola_jazz2021}.

Beyond infrastructural barriers, the feasibility of community-aware features also depends on how public service chatbots navigate organizational boundaries and ethical constraints. As Star and Griesemer note, boundary objects facilitate coordination across heterogeneous social groups \cite{star1989institutional}. In the context of community-level civic chatbots, boundary objects entail carefully defining who can participate in collective reporting and action, while safeguarding against disruptive or inauthentic engagement. The boundary of ``community'' must extend beyond geography to incorporate both organizational constraints (e.g., jurisdiction and agency responsibility) and social cohesion (e.g., shared values or local interests) \cite{star1989institutional, leigh2010not}. At the same time, visibility boundaries must also be calibrated, as transparency at the community level may conflict with individual privacy, especially when reporting sensitive civic issues \cite{becker_privacy_2019}. Drawing on the concept of organizational knowledge spaces \cite{erickson2000social}, visibility boundaries should be dynamically configured across different stakeholder groups---such as community members, local leaders, and government officials---to support accountability while preserving privacy. How, when, and to whom collective signals (e.g., co-reports, community sentiment, escalation status) should be made visible remains an open question at the intersection of interface design, institutional governance, and civic trust.

Being social-oriented alone is not inherently beneficial. The proposed community-level features must be critically examined for potential exclusions, power imbalances, and unintended consequences. Community-oriented chatbots may marginalize certain groups of people. \citet{pak2017fixmystreet} stated that digital participation platforms tend to marginalize low-income and ethnically diverse communities, with potential explanations including disparities in technical aptitude, language proficiency, and access to technology. Certain groups may dominate the engagement process, while others, such as marginalized communities, may be underrepresented, leading to decisions that do not reflect the entire community's needs \cite{abbott1995community, pak2017fixmystreet}. Furthermore, more community participation does not always mean more influential outcomes, as external stakeholders, such as funding agencies or governmental agencies, often hold the ultimate decision-making power \cite{khwaja2004increasing, baccaro2009downside}. These structural limitations highlight the need for critical reflection: simply adding participation mechanisms is not enough, and ensuring that engagement leads to meaningful and representative outcomes remains a complex, open design challenge.

\section{Conclusion}
This study examined the design and impact of a 311 chatbot in a metropolitan city on public service delivery and civic engagement. By combining official open-source survey data with 16 interpersonal interviews, we identified two challenge areas for public service chatbots: individual-level and community-level concerns. On the individual level, there are two key challenges: interpretation, transparency, and social contextualization. Moreover, the current chatbot design prioritizes the efficient completion of individual tasks but neglects the broader community perspective. To design from the perspective of the community, we highlighted three design opportunities: inform, report, and act collectively. Adding these social-oriented functions to the previous task-oriented chatbots requires customized proactive functions, which is location-based, case-driven, and context-aware. Designing at the community level raises challenges, including privacy, bias, and community divergence, which need to be explored by future research.

\bibliographystyle{ACM-Reference-Format}
\bibliography{sources.bib}
\end{document}